\documentclass[aps,prb,preprint,groupedaddress,longbibliography]{revtex4-1}
\usepackage{amsmath,amssymb,amsfonts,graphicx,times}
\usepackage{bbold}
\usepackage{graphicx}
\usepackage{dcolumn}
\usepackage{bm}
\usepackage{color}

\begin{document}


\title{Can a quantum critical state represent a blackbody?}


\author{Sudip Chakravarty}
\email{sudip@physics.ucla.edu}
\author{Per Kraus}
\email{pkraus@ucla.edu}
\affiliation{Mani L. Bhaumik Institute for Theoretical Physics\\ Department of Physics and Astronomy\\ University of California Los Angeles, Los Angeles, CA 9095, USA}


\date{\today}

\begin{abstract}
The blackbody theory of Planck played a seminal role in the development of quantum theory
at the turn of the past century.  A blackbody
cavity is generally thought to be a collection of photons in thermal equilibrium; the radiation emitted is at all wavelengths, and the intensity follows a scaling law, which is  Planck's characteristic distribution law. These photons arise from non-interacting normal modes. Here we suggest that certain
quantum critical states  when heated emit ``radiation'' at all wavelengths and satisfy all the criteria of a blackbody. An important difference is that the ``radiation'' does not necessarily consist of non-interacting photons, but also emergent relativistic bosons or fermions. The examples we provide  include emergent relativistic fermions at a topological quantum critical point. This  perspective on a quantum critical state may  be illuminating in many unforeseen ways.

\end{abstract}


\maketitle
\section{Introduction}
A significant   discovery of Planck was the fundamental constant of Nature, $h$ ($\hbar=h/2\pi$), which bears his name. The blackbody  distribution of photons has survived significant tests in nature, although the idea of quantum discreteness of energy has a more complex history.~\cite{Kuhn:1987} Two most important observations were a scaling law and the consistency with the Stefan-Boltmann law. Today we consider a black hole as a blackbody emitting Hawking radiation. In fact, it is also strongly argued, and experimentally determined from the cosmic microwave background, that the universe is a nearly perfect blackbody bathed in a radiation  at a temperature of 2.7 $K$.

One might like to raise the question as to if there are other instances of blackbody radiation that may be interesting to study. We give several examples of such  a possibility in condensed matter systems involving quantum critical points (QCP) and show how scale and conformal invariance play an  important role in this matter. At the very outset we would like to dispel a  possible misunderstanding. First, when we say radiation from a quantum critical point, we mean radiation  from a state of matter tuned to a quantum critical point, {\em not radiation from a single point in the phase space}. Second, massless relativistic fermions can equally well provide a bonafide example of blackbody ``radiation'' (from hereon we shall omit the quotation mark, unless there is any possible confusion).  However, what we have here are emergent relativistic fermions at topological quantum critical points tuned by the chemical potential,~\cite{Goswami:2011,Goswami:2016} not simply {\em ad hoc} non-interacting theories of free fermions.

More ambitious questions regarding what can perhaps be termed as interacting non-fermi liquid QCPs are reserved for the future. This is a thorny question: what appears to be strongly interacting Hamiltonian, under some clever choices of degrees of freedom, may represent noninteracting degrees of freedom. An example is an Ising model in a transverse field (TFIM)  in $(1+1)$-dimensions, which by Jordan-Wigner transformation can be cast into a  spinless free fermion theory with zero chemical potential at the QCP, separating a quantum disordered state from a spontaneously broken ferromagnetic state.~\cite{Pfeuty:1970} These fermions are  nonlocal in character, however. On the other hand, in the original spin variables TFIM is a strongly interacting problem with anomalous scaling dimensions. In the same spirit, we only consider those systems that can be transformed to noninteracting systems, however strongly interacting the original Hamiltonian may be. It should be kept in mind that there is a QCP separating two states of matter, and is thus not a trivial problem by any means.

Two important aspects of Planck's theory are worth focusing on. The first is the Stefan-Boltzmann law and the second is a scaling function that forms the basis of Wien's displacement law. The first  states that the energy density of radiation, $u\propto T^{d+1}$ in $d$-dimensions; more generally  $u\propto T^{\frac{d}{z}+1}$ to be discussed later, where $z$ is the dynamical critical exponent reflecting the anisotropy of scaling of time and space; the amplitude can contain additional physics  (such as the central  charge  in a conformal field theory). The second is the phenomenology  of the Wien's law. In three spatial dimensions, the energy density per unit wavelength,  $u_{\lambda}$, is
\begin{equation}
u_{\lambda}=\frac{c^{4}}{\lambda^{5}}F(\frac{c}{\lambda T}),
\end{equation}
where $c$  is the velocity of light and $\lambda$ is the wavelength of radiation,  and we have set the Planck and Boltzmann constants to unity.  Here $F$ is a scaling function. In electrodynamic theory of non-interacting photons the frequency is uniquely related by $ \omega = c k$. For later reference, let us rewrite it as
\begin{equation}
u_{\lambda}=\frac{T^{4}}{\lambda}G(\frac{c}{\lambda T}),
\end{equation}
where $G(\frac{c}{\lambda T})=(\frac{c}{\lambda T})^{4}F(\frac{c}{\lambda T})$ is another scaling function with the same argument.
The radiation  exists at all all length scales satisfying the scaling law.

\section{What is a quantum critical point?}
In this section we will provide a lightning summary of quantum critical points, at least those aspects of it that are relevant
for the present discussion. Quantum criticality is a concept pertinent to zero temperature ($T=0$).~\cite{Hertz:1976} A tuning parameter
can drive a complex many-body system to a point $g_{c}$ (a generic coupling constant for the time being), where quantum fluctuations exist at all length scales, from the lattice scale
to the correlation length $\xi=\infty$. But we cannot directly observe these remarkable fluctuations, because all experiments are necessarily carried out at a non-zero $T$. It is only through its influence on finite temperature observables that we can infer
this phenomenon.~\cite{Chakravarty:1988,*Chakravarty:1989} There are now very good arguments and experiments  that show that when tuned to $g_{c}$, the quantum criticality can extend to temperatures as large as the dominant  fundamental energy scale of the Hamiltonian.~\cite{Kopp:2005,Imai:2014}

Similar to blackbody radiation we can define a scaling function (spectral function) at $T\ne 0$ by
\begin{equation}
A(k,\omega,g, T)= L_{\tau}^{y_{A}}A(k L_{\tau}^{1/z},\omega L_{\tau}, L_{\tau}/\xi_{\tau})
\end{equation}
where $\omega$  the frequency, and  $k$ the wave vector, are two independent variables and are not necessarily tied to each other, as in the case of a photons; $y_{A}=d_{A}z$, $d_{A}$ is the scaling dimension of the operator $A$. The important scales are $L_{\tau}=\hbar/k_{B}T$ and the correlation timescale in imaginary time, $\tau$, given by  $\xi_{\tau}\sim \xi^{z}$. The spatial correlation length $\xi \sim (|g-g_{c}|/g_{c})^{-\nu}$. Thus, $\xi_{\tau}^{1/z}$  defines the spatial length scale $\xi$. Here $z$, $\nu$ and $y_{A}$ are three independent exponents.

At $T=0$, a bit of care is needed to define the dynamic scaling function,
\begin{equation}
A(k,\omega,g) = \xi^{y_{A}} A (k\xi,\omega \xi_{\tau}).
\end{equation}
 However, when tuned exactly to $g_{c}$, the quantum critical point at $T=0$
 \begin{equation}
A(k,\omega,g_{c}) = k^{-y_{A}} A (k^{z}/\omega).
\end{equation}
This is because at $T=0$ both $\xi_{\tau}$ and $\xi$ are infinite, so the only frequency scale left is $k^{z}\sim \omega$.

Let us now return to $T\ne 0$ but tuned to the $T=0$ quantum critical point $g_{c}$. Then a simple rearrangement leads to
\begin{equation}
A(k,\omega,g_{c}, T) =\left({\frac{1}{T}}\right)^{y_{A}}\widetilde A\left(\frac{k^{z}}{T},\frac{\omega}{T}\right),
\end{equation}
where we have set $\hbar=k_{B}=1$. This is the most general form of the displacement law. If we set $z=1$, with slight abuse of notation
we can write,
\begin{equation}
A(k,\omega,g_{c}, T) =\left({\frac{1}{T}}\right)^{y_{A}}\widetilde A\left(\frac{v k}{T}\right),
\end{equation}
where $\omega=vk$, $v$ is an excitation velocity, which will be made more explicit on a case by case basis.

In the simplest possible scenario of a  quantum critical point at $T=0$, there is one relevant parameter $g$ such that for $g>g_{c}$ the system flows to an attractive fixed point, defining a phase of matter with zero correlation length (well, almost), as  we coarse grain the system, while for $g<g_{c}$, it flows to another phase. At the repulsive fixed point $g=g_{c}$, there are no flows and the correlation length is infinity, scale invariant. There are fluctuations on all length scales and time scales as discussed above. This flow is defined in the language of a differential equation in terms a dimensionless length scale:
\begin{equation}
\frac{d g}{d\ln \ell} = \beta({g}),
\end{equation}
thus defining the renormalization group $\beta$-function.

\section{Conformal and scale invariance}

QCPs are described by scale invariant quantum field theories.  In essentially all known cases of physical relevance, QCPs with dynamical critical exponent $z=1$ have a traceless stress tensor, implying that scale invariance is promoted to conformal invariance,~\cite{Francesco:1997} and we assume this in mostly what follows. Scale invariance alone permits the trace of the stress tensor to be equal to the divergence of a local operator, while conformal invariance requires a strictly vanishing trace.  In one spatial dimension the former implies the latter, while proving this in higher dimensions remains an outstanding problem.  See Ref.~\onlinecite{Nakayama:2015} for a review.  For $z\neq 1$ tracelessness of the stress tensor is replaced by a more general relation discussed in the Appendix.

Let us consider a theory in the vicinity of a QCP, as described by the action $S = S_{\rm QCP} +   \int d^{d}x \sum_{i} g_{i}{\cal O}_{i}$, and for the moment consider simply $z=1$.
%
Then the stress tensor obeys the trace relation
\begin{equation}
T^{\mu}_{\mu} = \sum_{i}\Delta_{i}g_{i}{\cal O}_{i},
\end{equation}
Here $\Delta_{i}=d_{i}+\gamma_{i}$ is the scaling dimension of ${\cal O}_i$, expressed in terms of the ``engineering dimension" $d_i$ and the anomalous dimension $\gamma_i$
. A common case is where one has a  classically scale invariant theory, so all $d_{i}$ vanish. We
then usually write $\beta_{i}= \frac{dg_{i}}{d\ln \ell}= \gamma_{i}g_{i}$ and so
\begin{equation}
T_{\mu}^{\mu} = \sum_{i}\beta_{i}(g_{i}){\cal O}_{i}
\end{equation}
For simplicity, if we consider only one operator ${\cal O}$, we have a remarkable identity
\begin{equation}
T_{\mu}^{\mu} = \beta({g}) {\cal O}
\end{equation}
This formula can be used in both directions: conformal invariance implies $T_{\mu}^{\mu}=0$ and so does the vanishing of $\beta(g)$ at the quantum critical point at $g=g_{c}$ at $T=0$.~\cite{Cardy:2010}

If we confine ourselves to $g_{c}$ and increase the temperature, we can apply thermodynamic arguments to deduce the famous Stefan-Boltzmann law in any dimension up to a constant that cannot be deduced from thermodynamics alone. The calculation is well-known and elementary. We assume that the radiation emitted leads to a pressure $T_{\mu}^{\mu}=P-u/d=0$, where $u$ is the energy density. Because $T$ is traceless, when tuned to $g_{c}$ in $d$-dimensional space of volume $V=L^{d}$, the total energy $E=uV$ will obey
the thermodynamic relation,
\begin{equation}
\left(\frac{\partial E}{\partial V}\right)_{T}= T\left(\frac{\partial P}{\partial T}\right)_{V} - P
\end{equation}
It immediately follows that
\begin{equation}
(d+1)\frac{dT}{T} =\frac{du}{u},
\end{equation}
hence
\begin{equation}
u\propto T^{d+1}
\end{equation}
which is the Stefan-Boltzmann law. The proportionality constant hides crucially important physics, which is where the central charge  enters.
This is energy density {\em not} the power emitted and thus non-vanishing even in one dimension.

\section{Transverse field Ising model as a nontrivial example}

At a quantum critical point fluctuations of appropriate degrees of freedom diverge. However, what constitutes appropriate degrees of freedom is an interesting question. We will try to elaborate on this question by an explicit and simple (or not so simple) example of the one-dimensional
 transverse field Ising model  (with a beautiful experimental realization~\cite{Imai:2014}),  whose connection with the blackbody radiation can be illustrated.

\begin{figure}
 \includegraphics[width=0.8\textwidth]{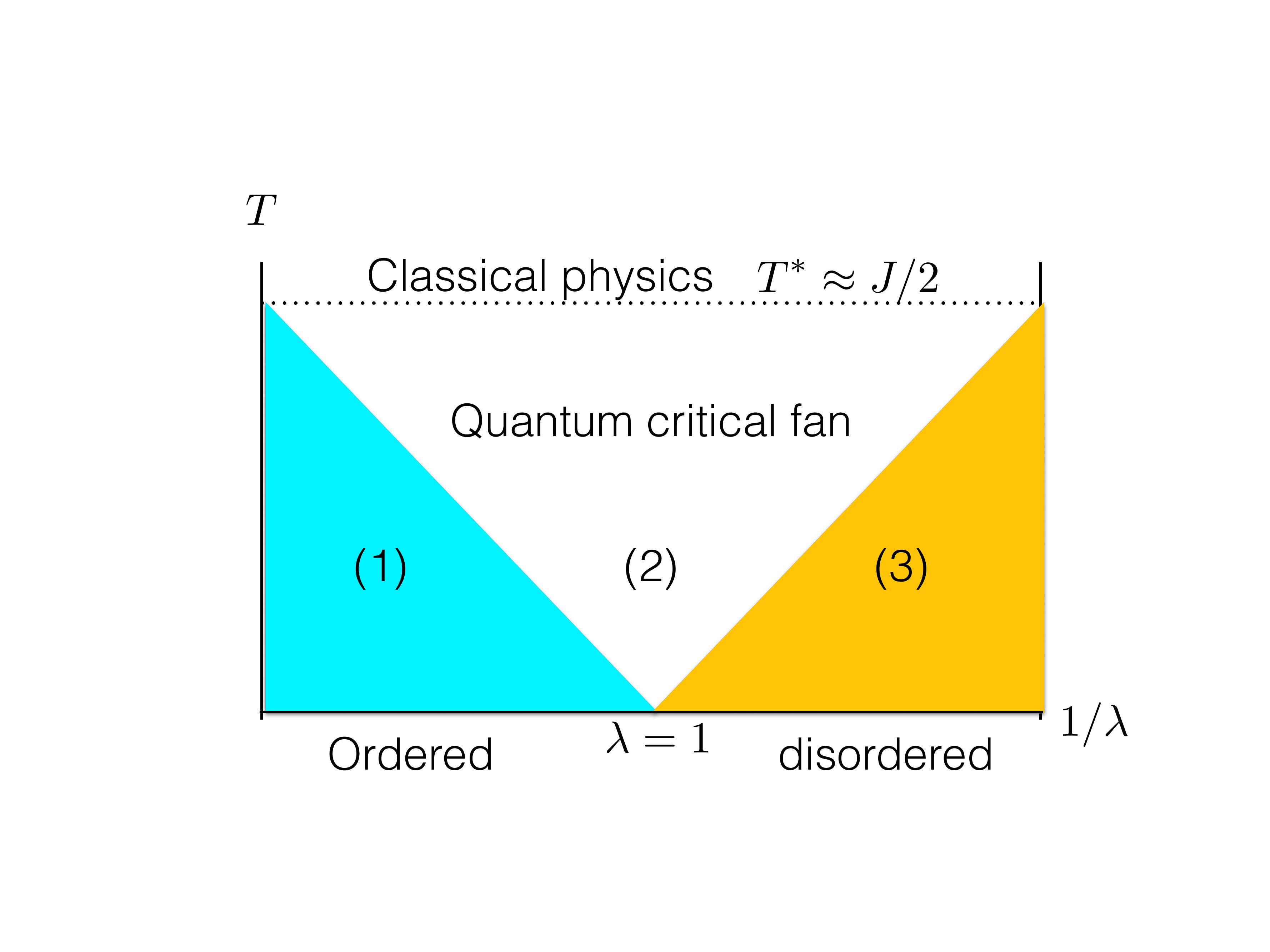}%
 \caption{Phase diagram for TFIM. The region $(3)$ is the quantum disordered state, $(2)$ is the quantum critical regime, and (1) is the renormalized classical regime in the terminology of Chakravarty, Halperin and Nelson.~\cite{Chakravarty:1988,*Chakravarty:1989} The temperature $T^{*}=J/2$ determines the scale up to which quantum criticality effectively persists.~\cite{Kopp:2005}\label{TFIM}}
 \end{figure}

The Hamiltonian of TFIM is
\begin{equation}
{\cal H}= - h \sum_{i}\sigma_{i}^{x} - J\sum_{i}\sigma_{i}^{z}\sigma_{i+1}^{z},
\end{equation}
where the $\sigma$'s are the conventional Pauli matrices. An exact result is that the critical point is at $\lambda=h/J=1$. For $\lambda>1$ the system is quantum disordered, a paramagnet at $T=0$, and for $\lambda < 1$ it is a ferromagnet with  spontaneously broken $\mathbb{Z_{2}}$ symmetry. The phase diagram is shown in Fig.~\ref{TFIM}.
By the well-known Jordan-Wigner transformation~\cite{Pfeuty:1970}, this Hamiltonian can be diagonalized in terms of  free (but non-local) spin-less fermions, as
\begin{equation}
{\cal H} = \sum_{k} \varepsilon_{k}\left(c^{\dagger}_{k}c_{k} - \frac{1}{2}\right),
\end{equation}
where
\begin{equation}
\varepsilon_{k} = 2J\left(1+\lambda^{2}-2\lambda \cos k\right)^{1/2};
\end{equation}
when linearized around the quantum critical point  the dispersion relation at large wavelengths is
\begin{equation}
\varepsilon_{k}\approx 2J\sqrt{2}\left|\sin\frac{k}{2}\right|\approx  \sqrt{2}J |k|
\end{equation}
The energy density at  low temperature $T$ is then to the leading approximation
\begin{eqnarray}
u_{F} &\approx& \frac{1}{2\pi}\int_{-\infty}^{\infty}dk\frac{\hbar v |k|}{e^{\hbar v|k|/k_{B}T} +1}\nonumber\\
&=&\frac{\pi}{12}\frac{T^{2}}{\hbar v}=\frac{c\pi}{6\hbar v}T^{2}
\end{eqnarray}
where the velocity $v=\sqrt{2} J$, and buried in this expression is the  central charge $c=1/2$ for spinless fermions.
We can repeat the same calculation for free relativistic bosons. Then,
\begin{equation}
u_{B} = \frac{\pi}{6}\frac{T^{2}}{\hbar v_{B}},
\end{equation}
where $v_{B}$ is the velocity of bosons. The central charge for bosons is unity. In either case one cannot tell what the degrees of freedom  are without scrutinizing carefully the prefactor.
\begin{figure}[htb]
 \includegraphics[width=0.6\textwidth]{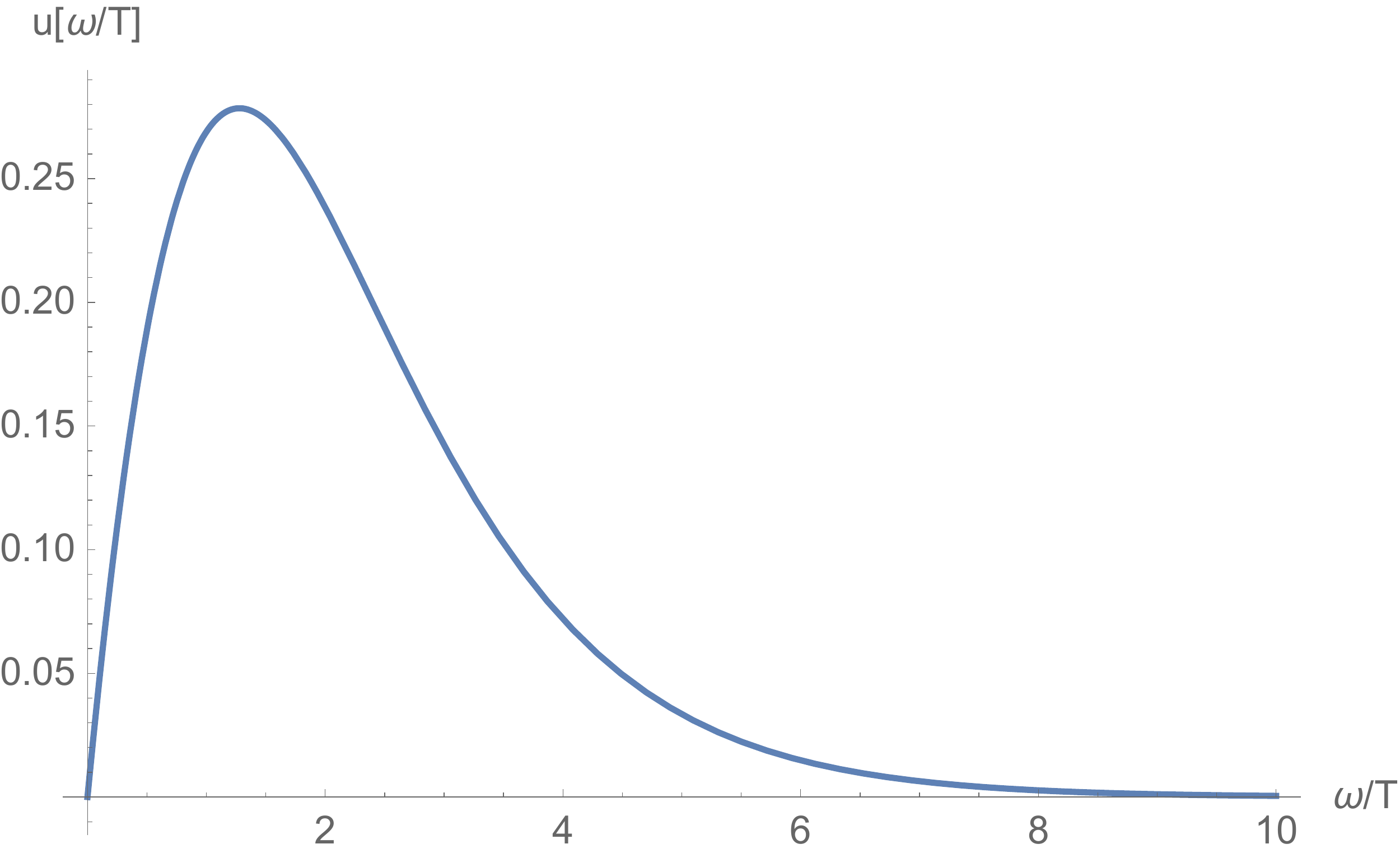}
 \caption{Energy density for Jordan-Wigner fermions for TFIM, $\omega=vk$,\label{JW}}
 \end{figure}
The Wien displacement law follows trivially. Nowhere does the anomalous dimension of the respective fermion or boson operators enter. One could equally well deduce the results for the temperature dependence from thermodynamics.

\begin{figure}[h]
 \includegraphics[width=0.6\textwidth]{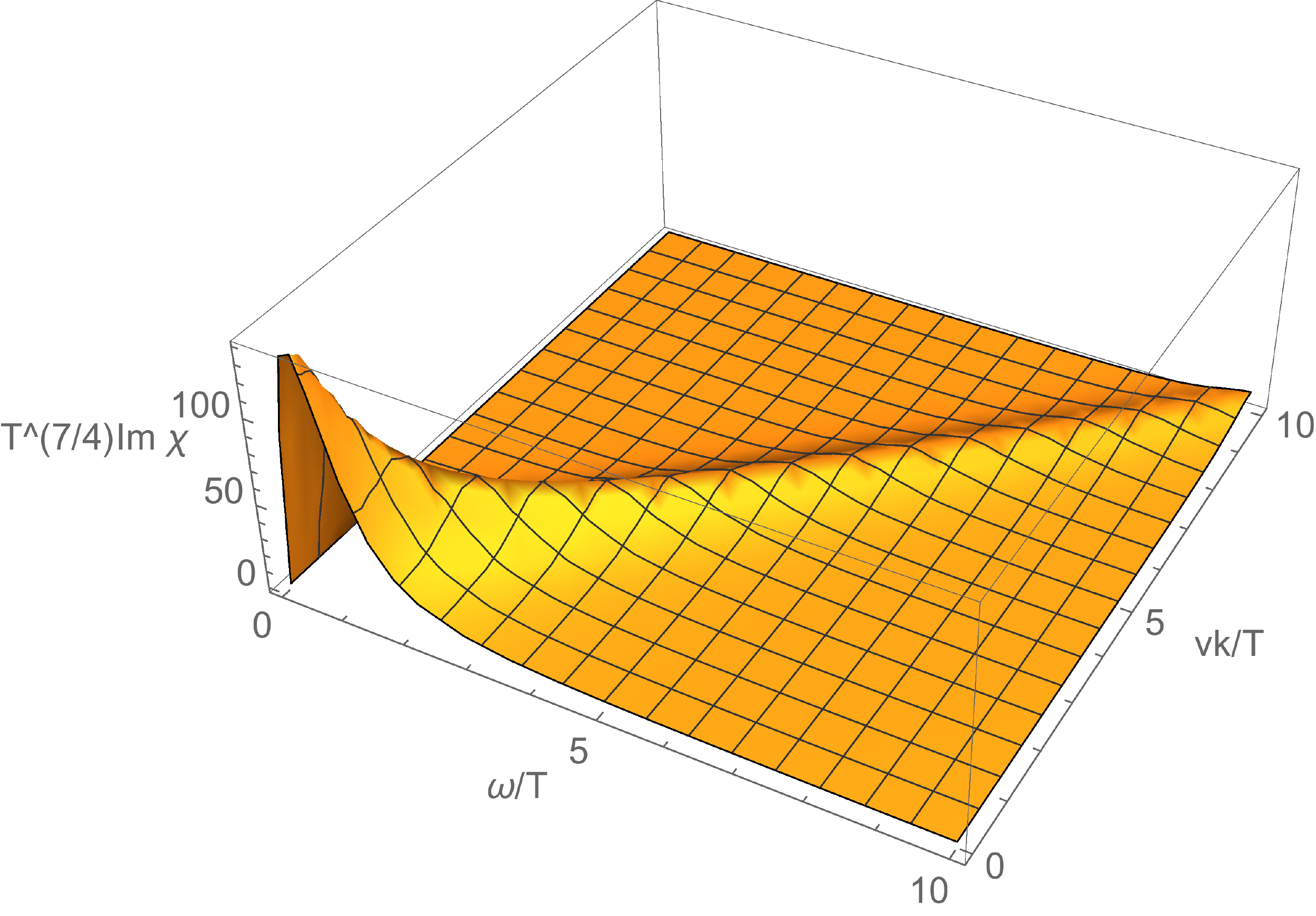}
 \caption{Imaginary part of the spin suceptibility for TFIM \label{Suscp}}
 \end{figure}

In terms of the original spin variables, Jordan-Wigner fermions are non-local objects, as is well known:
\begin{eqnarray}
c_{i}&=& \left(\prod_{j<i}\sigma_{j}^{z}\right)\sigma_{i}^{+}, \\
c_{i}^{\dagger}&=&\left(\prod_{j<i}\sigma_{j}^{z}\right)\sigma_{i}^{-}.
\end{eqnarray}
The inverse is also non-local.
\begin{eqnarray}
\sigma_{i}^{+}&=& \prod_{j<i}(1-2c_{j}^{\dagger}c_{j})c_{i}, \\
\sigma_{i}^{-}&=& \prod_{j<i}(1-2c_{j}^{\dagger}c_{j})c_{i}^{\dagger}.
\end{eqnarray}
We do not see how one can locally couple to a single Jordan-Wigner fermion; so direct verification of the blackbody spectrum is probably not possible.
On the other hand, the correlation function of $\sigma_{z}$ can be measured in neutron scattering from the frequency and momentum dependent susceptibility $\chi(k,\omega)$, which when tuned to criticality is~\cite{Schulz:1986}
\begin{equation}
\chi(k,\omega)\propto \frac{1}{T^{7/4}}\frac{\Gamma(\frac{1}{16} - i \frac{\omega+vk}{4\pi T})\Gamma(\frac{1}{16} - i \frac{\omega-vk}{4\pi T})}
{\Gamma(\frac{15}{16} - i \frac{\omega+vk}{4\pi T})\Gamma(\frac{15}{16} - i \frac{\omega-vk}{4\pi T})}
\end{equation}
The imaginary part of $\chi(k,\omega)$ gives the fluctuation spectra shown in Fig.~\ref{Suscp}. One can clearly extract the characteristic  velocity $v$, which is simply related to the exchange constant in TFIM.
 \section{Higher Dimensional Models}
We now provide examples of two higher dimensional models, where  the QCP is controlled by massless Dirac fermions.
Given the recently demonstrated idea of superuniversality,
one can provide many more examples,~\cite{Goswami:2016} but we hope that these will suffice to make our point.
\subsection{(2+1)-dimensions}
First consider the low energy Hamiltonian of spinless fermions  in two dimensions~\cite{Read:2000} with the spinor $\Psi_{k}^{\dagger}=(c_{k}^{*}, c_{-k}$):
\begin{eqnarray}\label{3dH0}
H=\frac{1}{2}\int \frac{d^2k}{(2\pi)^2} \; \Psi^\dagger_{\mathbf{k}} \widehat{{\cal H}}_{\mathbf{k}} \Psi_{\mathbf{k}},
\end{eqnarray}
where
\begin{equation}
\widehat{{\cal H}}_{k}= \hbar v (k_{x}\sigma_{x}+k_{y}\sigma_{y})+(-\mu+\hbar^{2}k^{2}/2m^{*})\sigma_{z}.
\end{equation}
Here $\sigma_{x}$, $\sigma_{y}$, and $\sigma_{z}$ are the conventional Pauli matrices. Dirac fermions have a momentum dependent mass $m_{\mathbf{k}}=-\mu + \hbar^2k^2/(2m^\ast)$ with $\mu$ being the chemical potential; the excitation velocity $v=\Delta_t/\sqrt{2 m^\ast |\mu|}$ with $\Delta_t$ the  triplet pairing amplitude.
The energy dispersion is
\begin{equation}
E_{k}=\pm\sqrt{\hbar^{2}v^{2}k^{2}+m_{k}^{2}},
\end{equation}
and the Hamiltonian can be brought  to the  form
\begin{equation}
{\cal H}_{k}= |E_{k}| \hat{n}\cdot \hat{\sigma}.
\end{equation}
The unit vector $\hat n$ is
\begin{equation}
\left(\frac{\hbar v k_{x}}{|E_{k}|}, \frac{\hbar v k_{y}}{|E_{k}|}, \frac{m_{k}}{|E_{k}|}\right)
\end{equation}
At $k=0$,
\begin{equation}
\hat{n} =\left(0,0,\frac{-\mu}{|\mu|}\right)
\end{equation}
and at $k\to \infty$
\begin{equation}
\hat{n} =\left(0,0,1\right).
\end{equation}
Therefore, when $\mu>0$, there is a skyrmion with wrapping number unity in the BCS (Bardeen-Cooper-Schrieffer) phase. This corresponds to homotopy
$\Pi^{2}(S^{2}) = \mathbb{Z}$. On the other hand for $\mu <0$, the BEC (Bose-Einstein Condensation) phase does not admit  skyrmions. Consequently, the topological distinction between the BEC and BCS states arises through the sign of the uniform Dirac mass or the chemical potential of the normal quasiparticles. Note that time reversal symmetry is broken and at QCP ($\mu=0$) the excitations are simply massless Dirac fermions in the low energy limit. Consequently,
one is led to a simple blackbody radiation as the temperature is turned on. The physical context could be $p_{x}+ip_{y}$ superconductivity in $\mathrm{SrRu_{2}O_{4}}$.~\cite{Kallin:2016} Note that this is an emergent low energy Hamiltonian, where superconductivity is described in terms of Bogoliubov-de Gennes theory. The massless Dirac spectrum emerges only when the system is tuned to the quantum critical point at $\mu =0$. The derivation of the properties of the blackbody is a simple  exercise.

How about  interactions and/or disorder effects? In a sense interactions are already present in forming the superconducting state, for example, a large negative $U$ Hubbard model to form the diatomic molecule picture of BEC. However, additional short range interactions and   mass disorder could also be added at the quantum critical point.  The four fermion interaction has the scaling dimension $(z-d)=-1$ and is therefore irrelevant. Similarly the scaling dimension of disorder is $(2z-d)=0$, which is marginal, but is actually known to be marginally irrelevant.~\cite{Evers:2008} Therefore, as long as we are aimed at the quantum critical point, the  fermionic  excitations remain valid and all the characteristics of blackbody radiation are satisfied.

\subsection{$(3+1)$-dimensions}
An analogous problem in the context of superfluid $\mathrm{^{3}He}$-B phase is given by the following low energy Hamiltonian~\cite{Goswami:2016}
\begin{eqnarray}\label{3dH0}
H=\frac{1}{2}\int \frac{d^3k}{(2\pi)^3} \; \Psi^\dagger_{\mathbf{k}} \widehat{{\cal H}}(\mathbf{k}) \Psi_{\mathbf{k}},
\end{eqnarray}
where $\Psi^\dagger_{\mathbf{k}}=(c^\ast_{\mathbf{k},\uparrow}, c^\ast_{\mathbf{k},\downarrow},c_{-\mathbf{k},\uparrow},c_{-\mathbf{k},\downarrow})$ is the four component Nambu spinor,  and $c_{\mathbf{k},s}$ is the annihilation operator for a normal state quasiparticle or fermionic $^3$He atom with spin projection $s=\uparrow / \downarrow$. The operator $\widehat{{\cal H}}(\mathbf{k})=\sum_{j=1}^{4} \; n_j(\mathbf{k}) \Gamma_j$, where we have introduced a four component vector
\begin{equation}
\mathbf{n}(\mathbf{k})=(\hbar vk_x,\hbar vk_y, \hbar vk_z,-\mu + \hbar^2k^2/(2m^\ast)),
\end{equation} with $\mu$ and $m^\ast$ respectively being the chemical potential and the effective mass of the normal quasiparticles.
The velocity $v=\Delta_t/(\hbar k_F)=\Delta_t/\sqrt{2 m^\ast |\mu|}$ with $\Delta_t$ being  the triplet pairing amplitude, and $\Gamma_j$ are four mutually anticommuting Dirac matrices: $\Gamma_1=-\sigma_3 \otimes \tau_1$, $\Gamma_2=-\sigma_0 \otimes \tau_2$, $\Gamma_3=\sigma_1 \otimes \tau_1$, $\Gamma_4=\sigma_0 \otimes \tau_3$. The Pauli matrices $\sigma_\mu$ and $\tau_\mu$ respectively operate on the spin and particle-hole  indices. By squaring the Hamiltonian one
can bring it to the form
\begin{equation}
\widehat{{\cal H}}= |E_{k}| \hat{n}\cdot \vec{\Gamma}
\end{equation}
where $\vec{\Gamma}$ is a four component vector composed of Dirac matrices defined above, and $\hat n$ now is a 4-component unit vector. Here $E_{k}= \pm\sqrt{\hbar^{2}v^{2}k^{2}+m_{k}^{2}}$,  $k^{2}=k_{x}^{2}+k_{y}^{2}+k_{z}^{2}$ and $m_{\mathbf{k}}=-\mu + \hbar^2k^2/(2m^\ast)$.
Once again, at $k=0$,
\begin{equation}
\hat{n} =\left(0,0,0,\frac{-\mu}{|\mu|}\right)
\end{equation}
and at $k\to \infty$
\begin{equation}
\hat{n} =\left(0,0,0,1\right).
\end{equation}
It is again an  example of a topological quantum criticality determined by the sign of $\mu$, leading to excitations consisting of $(3+1)$-dimensional Dirac fermions.
The skyrmion number is given by the standard expression
\begin{equation}
{\cal N}_{sk}=\frac{1}{12\pi^{2}}\epsilon_{ijk}\epsilon_{abcd}\int d^{3}k\; n_{a}\partial_{i}n_{b}\partial_{j}n_{c}\partial_{k}n_{d},
\end{equation}
corresponding to the homotopy $\Pi_{3}(S^{3})=\mathbb{Z}$. Consequently,
one is led to a simple blackbody radiation as the temperature is turned on, but now the time reversal invariance is respected. As before, additional short range interaction and   mass disorder could also be added at the quantum critical point.  A four fermion interaction has the scaling dimension $(z-d)=-2$ and  is therefore irrelevant. Similarly the scaling dimension of disorder is $(2z-d)=-1$, which is also irrelevant. Therefore, as long as we are aimed at the quantum critical point, the  characteristics of blackbody radiation are satisfied.
\subsection{Discussion}
So far we have considered examples  where the dynamical exponent $z=1$. When $z\ne1$, we must think anew.  We interpret the generalized scaling function $A$ to be the imaginary part of the retarded Green function, i.e.,
\begin{equation}
A(k,\omega,g_{c}, T) =\left({\frac{1}{T}}\right)^{y_{A}}\widetilde A\left(\frac{k^{z}}{T},\frac{\omega}{T}\right),
\end{equation}
The anisotropic stress-energy tensor  still satisfies an analogue of tracelessness and is proportional to the $\beta$-function as shown in the Appendix~\ref{ST}. Thus,  all the general characteristics of
a blackbody are satisfied with minor modifications, for example, the Stefan-Boltzman law is modified to $u\propto T^{\frac{d}{z}+1}$ (See also Ref.~\onlinecite {abrahams:2012}).

There is a great deal of interest in  non-Fermi liquids in condensed matter physics. They are mostly defined by power laws in transport coefficients, and these do not follow the Fermi liquid predictions. A famous example is the linear resistivity in high temperature superconductors, as a function of temperature, that extends over a wide range. But most importantly,  non-Fermi liquids  do not have quasiparticle poles in the spectral function but cuts.~\cite{Yin:1996}  This,  most likely,  could lead to differences with conventional  blackbody spectra. The other difference is that the spectra  will contain substantial  inhomogeneities due to the lack of  a quasiparticle description; the inhomogeneity in the cosmic microwave background has been measured, but it is very small. This topic will be discussed in the future. On the other hand, we cannot see how this could possibly change the robust thermodynamic properties when tuned to the QCP, as mentioned earlier.

 Unfortunately, the Jordan-Wigner fermions discussed earlier are highly non-local objects and cannot be measured by a local probe. On the other hand, if we could interpret them as the critical fermions at the BCS (Bardeen-Cooper-Schrieffer) to BEC (bose-Einstein Condensation) phase transition for a one-dimensional $p$-wave superconducting chain (which is an approximation in this case), they could be detected by a tunneling experiment. By contrast, consider the imaginary part of the spin susceptibility, as shown in Fig.~\ref{Suscp}, which reflects a strongly interacting system.
The distinction between the two  is quite striking. In terms of Jordan-Wigner fermions, the behavior is identical to that of a black body, peaking at an intermediate temperature for a given frequency. The maximum shifts  as in Wien's displacement law. On the other hand the spin susceptibility  diverges at  low frequencies, see Fig.~\ref{Suscp}. as it must since  $g$ is tuned to the quantum critical point $g_{c}$. The quantum critical behavior of TFIM has been beautifully detected in NMR experiments in weakly coupled Ising spin chains, known as Cobalt Niobate ($\mathrm{CoNb_{2}O_{6}}$).~\cite{Imai:2014} It is also possible to  measure the fluctuation spectrum in neutron scattering measurements.

Although non-Fermi liquids are implicated in a large number of materials, such as cuprate high temperature superconductors, heavy fermions, etc.,   the universality of the behavior calls for a more fundamental understanding;   the material dependence is of secondary importance. It is perhaps this goal that the present work may inspire. Of course, if  the specific heat at low temperature is measured and neutron scattering experiments  measure the characteristic excitation velocity, one can experimentally determine the central charge.  It is of course possible to tune away from $g_{c}$, but we wanted to present our work in the simplest possible case. Of course, most of our results are well known; we simply provided a new perspective to view them.
\begin{acknowledgments}
We are grateful to Pallab Goswami for many insightful comments. We would also like to thank Ching-Kit Chan, Steven Kivelson, and S. Raghu for discussions. The work was supported in part by funds from the David S. Saxon Presidential Term Chair at UCLA (S. C.).  Work of PK is supported in part by NSF grant 1619926.
\end{acknowledgments}
\appendix 
\section{Trace of the stress tensor}
\label{ST}

Here we review the relation between scale invariance and the trace of the stress tensor, including the case of anistropic scaling, $z\neq 1$.    See, for example, Ref.~\onlinecite{Baggio:2011}.

\subsection{Isotropic scaling: $z=1$}

We consider a general quantum field theory described by an action $S[\phi]$, and the corresponding path integral $Z=\int\! D\phi e^{-S[\phi]}$.   The theory at the critical point is assumed to have dynamical scaling exponent $z=1$, and lives on the flat Euclidean metric $g_{\mu\nu}=\delta_{\mu\nu}$.  We study the vicinity of the QCP by writing $S = S_{\rm QCP} + \int \sqrt{g} \sum_i g_i O_i$, where the operators $O_i$ have definite scaling dimensions at the QCP.  Let the coupling $g_i$ have classical mass dimension $d_i$.   Quantum mechanically we have to introduce an arbitrary renormalization scale $\mu$, and the couplings and operators depend on this scale. The renormalized path integral is written as $Z[g_{\mu\nu}, g_i(\mu),\mu]$.   Basic dimensional analysis tells us
\begin{equation}
Z[e^{2\sigma} g_{\mu\nu}, e^{-d_i\sigma} g_i(\mu),e^{-\sigma} \mu] = Z[g_{\mu\nu}, g_i(\mu),\mu]
\label{A2}
\end{equation}
Since the path integral doesn't depend on $\mu$ we have
\begin{equation}
Z[e^{2\sigma} g_{\mu\nu}, e^{-d_i\sigma} g_i(\mu),e^{-\sigma} \mu] = Z[e^{2\sigma} g_{\mu\nu}, e^{-d_i\sigma} g_i(e^{\sigma} \mu), \mu]
\end{equation}
and by definition of the anomalous dimension $\gamma_i$,
\begin{equation}
 g_i(e^\sigma \mu) = e^{-\gamma_i\sigma} g_i(\mu)
\end{equation}
The quantum statement of behavior under scale transformations is then
\begin{equation}
Z[e^{2\sigma}g_{\mu\nu},e^{-\Delta_i \sigma}g_i,\mu] = Z[g_{\mu\nu}, g_i(\mu),\mu]
\end{equation}
where $\Delta_i =d_i+\gamma_i$ is the full scaling dimension.   This implies the relation
\begin{equation}
2g_{\mu\nu}{\partial \ln Z \over \partial g_{\mu\nu} } =     \sum_i \Delta_i g_i {\partial \ln Z \over \partial g_i}
\end{equation}
We recall that the stress tensor is obtained by varying the metric,
\begin{equation}
\langle T_{\mu\nu}\rangle = -{2\over \sqrt{g}}{\delta \ln Z \over \delta g_{\mu\nu}}
\end{equation}
and so the trace is
\begin{equation}
\langle T^\mu_\mu\rangle = -{2\over \sqrt{g}} g_{\mu\nu} {\delta \ln Z\over \delta g_{\mu\nu}}
\end{equation}
Similarly, local operators insertions are obtained by varying the couplings,
\begin{equation}
\langle O_i\rangle= -{1\over \sqrt{g}}{\delta \ln Z \over \delta g_i}
\end{equation}
Combining these results, we arrive at the operator relation
\begin{equation}
 T^\mu_\mu = \sum_i \Delta_i g_i O_i
\end{equation}
up to total derivatives (whose possible presence is related to the possible existence of scale but not conformally invariant theories.)

A common case is where one has a classically scale invariant theory, so all $d_i$ vanish.   We then usually write $\beta_i =\mu {dg_i\over d\mu} =- \gamma_i g_i $ and so
\begin{equation}
 T^\mu_\mu = -\sum_i \beta_i O_i
\end{equation}

In the above we phrased the argument in terms of the renormalized path integral.
Alternatively, from a more standard condensed matter viewpoint one might instead work with an explicit UV cutoff $\Lambda$.  An analogous argument goes through.

\subsection{Anisotropic scaling: $z\neq 1$}

It's useful to think of the theory as being defined on a metric of the form
\begin{equation}
ds^2 = N^2 dt^2 + h_{ij}dx^idx^j
\end{equation}
with  a preferred time foliation.  For example, we can consider the following action with anisotropic scale invariance
\begin{equation}
S = \int\! dt d^{d-1}x N\sqrt{h}\left({1\over 2} N^{-2}(\partial_t \phi)^2 -{1\over 2}\phi (\nabla^i\nabla_i)^z \phi\right)
\end{equation}
Here $\nabla_i$ denotes the covariant derivative built out of $h_{ij}$.   For $z=1$ the action is Lorentz invariant (assuming constant $N$ and $h_{ij}$), but not otherwise.

Let us first discuss the classical theory.
We focus on scale transformations,
\begin{equation}
 h_{ij} \to  e^{2\sigma} h_{ij}~,\quad N \to e^{z\sigma} N~,\quad \phi \to e^{{z+1-d\over 2}\sigma} \phi
\end{equation}
under which the action is invariant.   When the equations of motion are satisfied the action is stationary, ${\delta S\over \delta \phi}=0$, and then we have
\begin{equation}
 2 h_{ij} {\delta S \over \delta h_{ij}}+zN{\delta S\over \delta N}=0
\end{equation}
We define the energy density $\cal E$ and spatial stress tensor (momentum flux) $\Pi^{ij}$ as
\begin{equation}
 {\cal E}= {1\over \sqrt{h}}{\delta S\over \delta N}~,\quad \Pi^{ij}= {2\over N \sqrt{h}}{\delta S \over \delta h_{ij}}
\end{equation}
so that  that
\begin{equation}
z{\cal E}+ \Pi^i_i=0
\end{equation}
which is the generalization of $T^\mu_\mu=0$ to theories with $z\neq 1$.

Turning to the quantum theory, we consider a scale invariant theory with path integral
\begin{equation}
Z[N,h_{ij}] = \int\! D\phi e^{-S}
\end{equation}
which obeys
\begin{equation}
 Z[e^{z\sigma} N, e^{2\sigma} h_{ij}]=Z[N,h_{ij}]
 \label{eq:inv}
\end{equation}
and so we have
\begin{equation}
\langle {\cal E} \rangle = -{1\over \sqrt{h}}{\delta \ln Z\over \delta N}~,\quad \langle \Pi^{ij}\rangle = -{2\over N\sqrt{h}} {\delta \ln Z\over \delta h_{ij}}
\end{equation}
This generalizes to include other operator insertions.    The invariance in Eq.~(\ref{eq:inv}) implies the operator equation $z{\cal E}+ \Pi^i_i =0$.

Now let's add other operators to the action
\begin{equation}
S = S_{\rm QCP}  + \int\! dt d^{d-1}x N\sqrt{h}\sum_i g_i O_i
\end{equation}
The couplings $g_i$ are taken to have engineering dimension $d_i$, by which we mean that classically
\begin{equation}
 Z[e^{z\sigma} N, e^{2\sigma} h_{ij},e^{-d_i\sigma}g_i]=Z[N,h_{ij},g_i]
\end{equation}
We also note
\begin{equation}
\langle O_i \rangle = -{1\over N\sqrt{h}}{\delta \ln Z \over \delta g_i}
\end{equation}

Now include quantum effects.  We need to introduce an arbitrary renormalization scale $\mu$, and the couplings $g_i(\mu)$ depend on this scale, as do the operators.   The previous  scale transformation must be accompanied by a scaling of $\mu$,
\begin{equation}
 Z[\lambda^z N, \lambda^2 h_{ij},\lambda^{-d_i}g_i(\mu),\lambda^{-1}\mu]=Z[N,h_{ij},g_i(\mu),\mu]
\end{equation}
On the other hand, the path integral is $\mu$ independent, so
\begin{equation}
Z[e^{z\sigma} N, e^{2\sigma} h_{ij},e^{-d_i\sigma}g_i(\mu),e^{-\sigma} \mu]=Z[e^{z\sigma} N,  e^{2\sigma} h_{ij},e^{-d_i\sigma}g_i(\lambda \mu),\mu]
\end{equation}
hence
\begin{equation}
Z[e^{z\sigma}N, e^{2\sigma}h_{ij},e^{-d_i\sigma}g_i(\lambda \mu),\mu]=Z[N,h_{ij},g_i(\mu),\mu]
\end{equation}
Using the definition of the $\beta$ function
\begin{equation}
 \mu {d\over d\mu} g_i(\mu) = \beta_{i}(g)
\end{equation}
we find
\begin{equation}
zN{\delta \ln Z\over \delta N} + 2h_{ij}{\delta \ln Z \over \delta h_{ij} } +(\beta_i-d_ig_i){\delta \ln Z \over \delta g_i}=0
\end{equation}
which implies the operator equation
\begin{equation}
z{\cal E} +\Pi^i_i = \sum_i (d_i g_i O_i -\beta_i O_i)
\end{equation}
Classically marginal operators have $d_i=0$.

\end{document}